\documentclass[switch]{article} % Method A for two-column formatting

\usepackage{preprint}

\usepackage{amsmath, amsthm, amssymb, amsfonts}

\usepackage[numbers,square]{natbib}
\usepackage[utf8]{inputenc}	% allow utf-8 input
\usepackage[T1]{fontenc}	% use 8-bit T1 fonts
\usepackage{xcolor}		% colors for hyperlinks
\usepackage[hidelinks]{hyperref}	% Color links to references, figures, etc.
\usepackage{booktabs} 		% professional-quality tables
\usepackage{nicefrac}		% compact symbols for 1/2, etc.
\usepackage{microtype}		% microtypography
\usepackage{lineno}		% Line numbers
\usepackage{float}			% Allows for figures within multicol
\usepackage{graphicx}
\usepackage{lipsum}		%  Filler text
\usepackage[bbgreekl]{mathbbol}
\usepackage{tabularx}
\usepackage{subcaption}
\usepackage{mathtools}
\DeclarePairedDelimiter\abs{\lvert}{\rvert}%
\DeclarePairedDelimiter\norm{\lVert}{\rVert}%
\makeatletter
\let\oldabs\abs
\def\abs{\@ifstar{\oldabs}{\oldabs*}}
\let\oldnorm\norm
\def\norm{\@ifstar{\oldnorm}{\oldnorm*}}
\makeatother

\usepackage[ruled,resetcount]{algorithm2e}
\makeatletter
\renewcommand*\l@algocf{\l@figure}
\makeatother
\SetAlgoInsideSkip{smallskip}
\SetAlCapNameFnt{\footnotesize}
\SetAlCapFnt{\footnotesize}

\graphicspath{{Figs/}}

\usepackage{newfloat}
\DeclareFloatingEnvironment[name={Supplementary Figure}]{suppfigure}
\usepackage{sidecap}
\sidecaptionvpos{figure}{c}

\usepackage{titlesec}
\titlespacing\section{0pt}{12pt plus 3pt minus 3pt}{1pt plus 1pt minus 1pt}
\titlespacing\subsection{0pt}{10pt plus 3pt minus 3pt}{1pt plus 1pt minus 1pt}
\titlespacing\subsubsection{0pt}{8pt plus 3pt minus 3pt}{1pt plus 1pt minus 1pt}

\title{Lookup tables for phase randomisation in hardware generated holograms}

\usepackage{authblk}

\author[1,*]{Peter J. Christopher}
\author[1]{Timothy D. Wilkinson}
\affil[1]{Centre of Molecular Materials, Photonics and Electronics, University of Cambridge, UK}
\affil[*]{pjc209@cam.ac.uk}

\begin{document}
    
    \maketitle
	
	\begin{abstract}
		The rise in virtual and mixed reality systems has prompted a resurgence of interest in two-dimensional and three-dimensional real-time computer generated holography. Phase randomisation is an integral part of holographic projection as it ensures independence in sub-frame techniques and reduces the edge enhancement seen in flat-phase images. Phase randomisation requires, however, the availability of a pseudo-random number generator as well as trigonometric functions such as $\cos$ and $\sin$. On embedded devices such as field programmable gate arrays and digital signal processors this can be an unacceptable load and necessitate the use of proprietary intellectual property cores. Lookup tables are able to reduce the computational load but can run to many megabytes for even low-resolution systems.
        
        This paper introduces the use of lookup tables (LUTs) in the context of two common algorithms used for real-time holographic projection: Gerchberg-Saxton and One-Step Phase-Retrieval. A simulated study is carried out to investigate the use of relatively small lookup tables where random numbers are repeated in sequence. We find that the increase in error is low and tunable to under $5\%$ even for small look up tables. This results is also demonstrated experimentally. Finally, the implications of this study are discussed and conclusions drawn.
	\end{abstract}
	
    \vspace{0.35cm}

    \normalsize
    %%%%%%%%%%%%%%%  Main text   %%%%%%%%%%%%%%%
    % \linenumbers
    %\twocolumn
    
	\section{Introduction}
	
    The decades since the first computer generated holograms (CGHs) have seen a rapid growth in their quality and application. Holograms are now widely used in technologies including displays~\cite{display1,Maimone2017}, projectors~\cite{Makowski2012,Makowski2018, Chang2017}, lithography~\cite{Turberfield2000}, telecommunications~\cite{Crossland2002}, beam shaping~\cite{Zhang2019}, imaging~\cite{Sheen2001, Daneshpanah2010} and optical tweezing~\cite{Grieve2009, Melville2003}. While early research focussed primarily on producing still frames for display on spatial light modulators~(SLMs), recent years have seen significant research in two newer areas, real-time video and three-dimensional (3D) displays~\cite{Nehmetallah2013}. Combined with the improvement in quality of available SLMs as well as an increase in user expectations of image quality have led to a computational problem orders of magnitude greater than that faced previously. Recent years have seen numerous efforts to tackle this including extensive use of Graphical Processing Units (GPUs), Field Programmable Gate Arrays (FPGAs) and Digital Signal Processors (DSPs).
    
    This paper seeks to present a method of improving the speed of hologram generation for two common hologram generation algorithms - Gerchberg-Saxton (GS) and One-Step Phase-Retrieval (OSPR) - in the context of embedded devices including FPGAs and DSPs. This is done by moving elements of the computation dependent on a random number generator to a look-up table (LUT). We show that this can remove the need for trigonometric functions and a pseudo random number generator (PRNG) from the embedded process at the expense of higher memory usase. We also show that significant reuse of the random data can occur in each frame before significant error is introduced. This is then demonstrated on an implementation of OSPR for a $1024\times1024$ binary ferroelectric display~\cite{freeman2010visor2}.
    	
	\begin{figure}[h]
		\centering
		{\includegraphics[trim={0 0 0 0},width=0.7\linewidth,page=1]{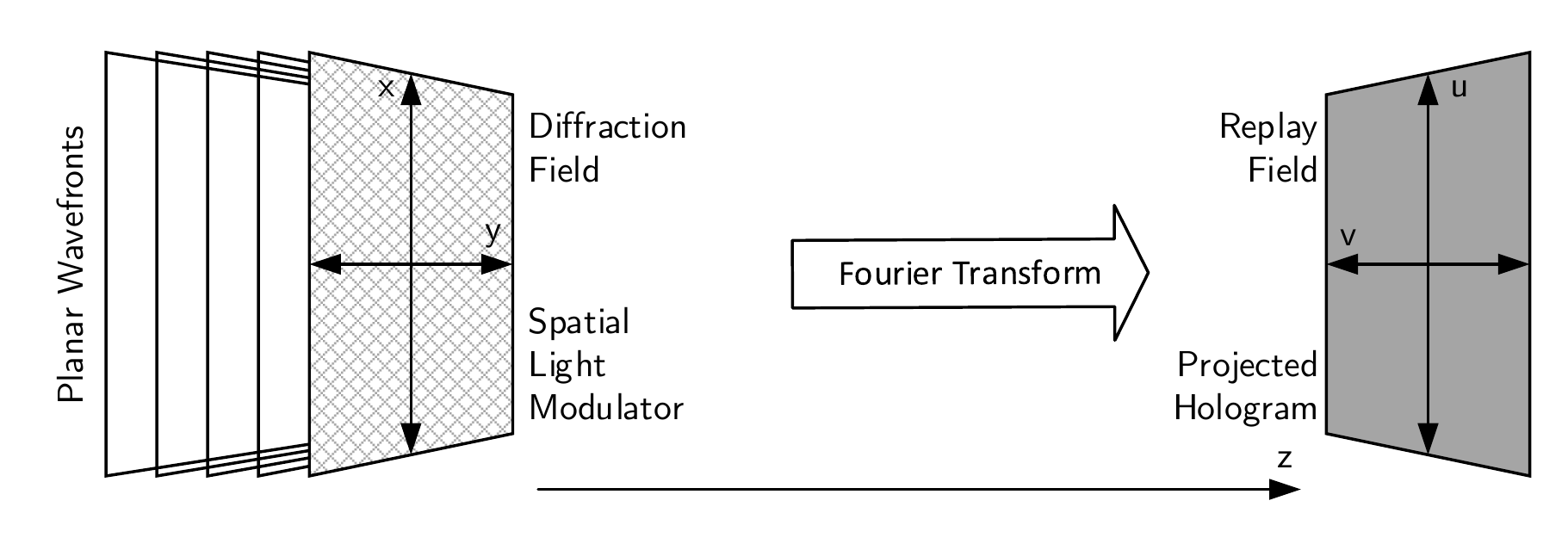}}
		\caption{Coordinate systems used in describing a hologram.}
		\label{fig:HoloCoordinateSystems}
	\end{figure}

	\section{Background}
	
	Figure~\ref{fig:HoloCoordinateSystems} shows the setup of a phase insensitive holographic system. A light field is passed through a SLM known as the diffraction field~(DF). The SLM applies a modulation or aperture function to the light that then propagates in free space before being scattered off of a diffuse screen. Depending on the ratio of propagation distance to SLM aperture size, this behaviour can be modelled as a Fourier or Fresnel transform. For constant illumination, planar wavefronts passed through a $100\%$ fill factor pixellated SLM, the far-field hologram displayed are given by the Discrete Fourier Transform~(DFT)~\cite{goodman2005introduction},
	
	\begin{align}
	F_{u,v} = \mathcal{F}\{f_{x,y}\}         & = \frac{1}{\sqrt{N_xN_y}}\sum_{x=0}^{N_x-1}\sum_{y=0}^{N_y-1} f_{xy}e^{-2\pi i \left(\frac{u x}{N_x} + \frac{v y}{N_y}\right)} \label{fouriertrans2d5c}   \\
	f_{x,y} = \mathcal{F}^{ - 1 }\{F_{u,v}\} & = \frac{1}{\sqrt{N_xN_y}}\sum_{u=0}^{N_x-1}\sum_{v=0}^{N_y-1} F_{uv}e^{2\pi i \left(\frac{u x}{N_x} + \frac{v y}{N_y}\right)}  \label{fouriertrans2d5d},
	\end{align}
	
	where $x$ and $y$ represent the source coordinates and $u$ and $v$ represent the spatial frequencies. Computational performance of $O(N_xN_y\log{N_xN_y})$ where $N_x$ and $N_y$ are the $x$ and $y$ respective resolutions can be achieved using the Fast Fourier Transform~(FFT) algorithm~\cite{Frigo2005,Carpenter2010}. Finding a given far-field hologram can be taken as the problem of finding discrete aperture function where $f(x,y)$ where $F(u,v) = \mathcal{F}\{f(x,y)\}$. 
	
	The majority of hologram generation algorithms begin by taking target complex replay field $R$ and back propagating it to the diffraction field $H=\mathcal{F}^{-1}\{R\}$ \cite{Park2017,tsang2016review}. The hologram is then modulated or \textit{quantised} to conform to the constraints of the SLM. Normally this is done by changing the value of each pixel $H_{x,y}$ to the nearest acceptable value $H'_{x,y}$. This is the \textit{initial hologram}. Different algorithms then take different approaches to changing pixels within the modulation constraints in order to reduce the error of the new replay field $R'=\mathcal{F}\{H'\}$.~\cite{MostChangedPixel}
	
\begin{figure}[h]
		\centering
		{\includegraphics[trim={0 0 0 0},width=0.3\linewidth,page=1]{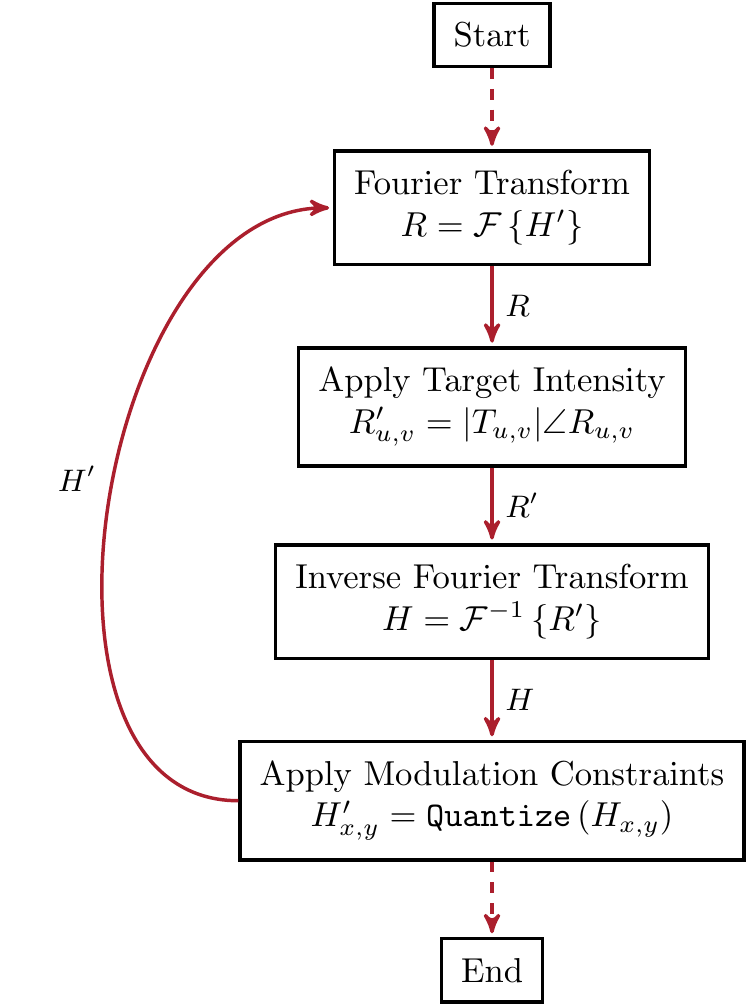}}
		\hspace{0.1\linewidth}
		{\includegraphics[trim={0 0 0 0},width=0.3\linewidth,page=1]{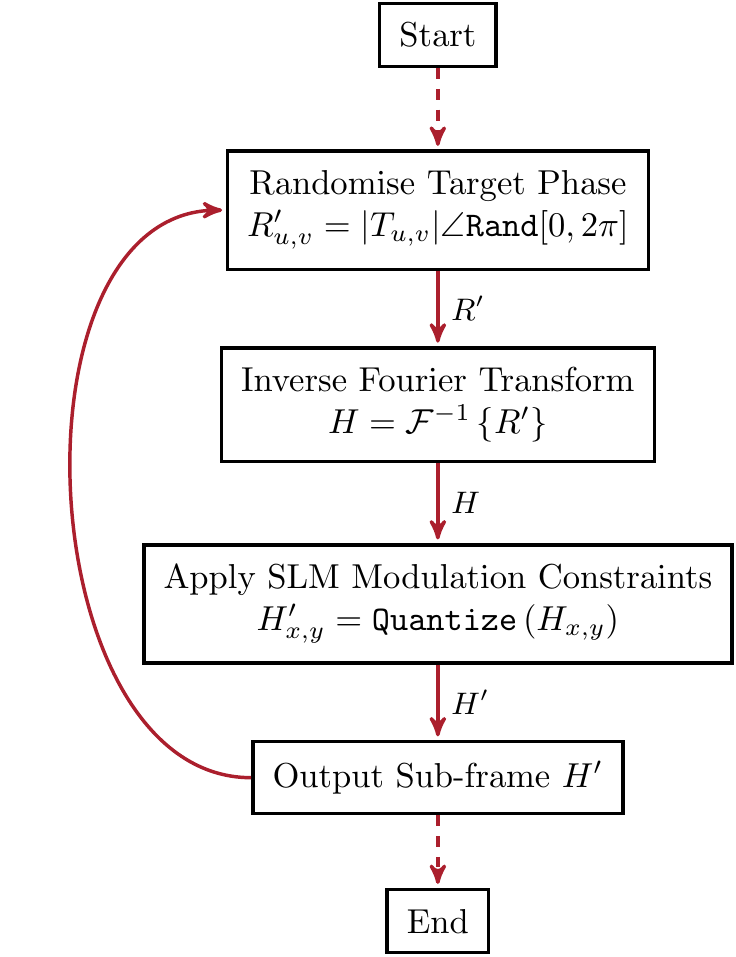}}
		\caption{Gerchberg-Saxton algorithm (left) and One-Step Phase-Retrieval algorithm (right).}
		\label{fig:algs}
	\end{figure}

	Many applications are phase insensitive and generation algorithms can focus solely on matching target amplitudes. In this case two algorithms dominate, Gerchberg-Saxton~(GS) and One-Step Phase-Retrieval~(OSPR) shown in Figures~\ref{fig:algs}~(left) and \ref{fig:algs}~(right) respectively~\cite{Gerchberg1972,Cable2004}. 
	
	\section{Phase randomisation}
	
	The phase randomisation shown in Figures~\ref{fig:algs}~(left) and \ref{fig:algs}~(right) for GS and OSPR is introduced for two primary reasons.
	
	Firstly, adding random phase on the target image will diffuse the object light, enabling not only the high frequency component, but also the low frequency component being recorded on the hologram plane. More information will be preserved on the hologram plane. This has the effect of preventing edge enhancement ~\cite{Shimobaba2015}.
	
	Secondly, OSPR introduces phase randomisation to ensure speckle independence of the each reconstructed target sub-frames~\cite{Maimone2017}. Time averaging then ensures a reduction in Mean Squared Error~(MSE). 

	\section{Motivation}
	
	Software implementations of these algorithms are able to use PRNGs~\cite{Shimobaba2015a} and trigonometric functions to carry out the phase randomisation. Field Programmable Gate Arrays~(FPGAs) or Digital Signal Processors~(DSPs) are also capable of this but can require the use of proprietary IP cores for the PRNG and trigonometric functions. To improve performance, Lookup Tables~(LUT) can be used. Random numbers can be generated at compile time to fill the LUT which then acts as a \textit{pool} of random numbers for algorithms to sequentially draw from. It the length of the random number LUT is given by $N_{\text{LUT}}$ then the case where $N_{\text{LUT}}$ is greater than the size of the test image is indistinguishable from the case without a LUT. LUTs of this size can be impractical on low-cost devices~\cite{Shimobaba2016}.
    
    \section{Results}
    
\begin{figure}[h]
        \centering
        {\includegraphics[width=0.5\linewidth,page=1]{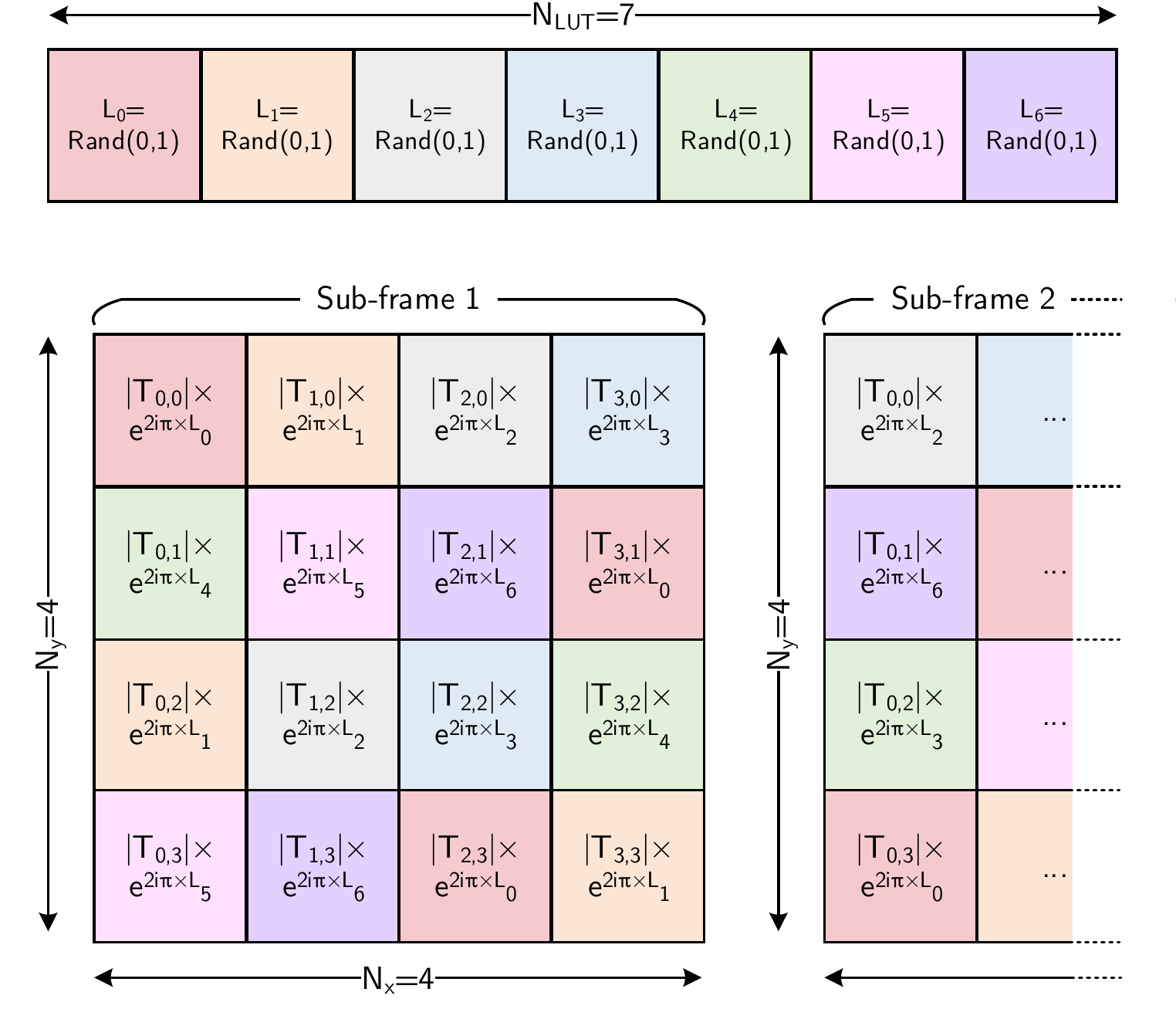}}
        \caption{Example of phase randomisation using look-up tables, each block represents a single pixel.}
        \label{fig:lut}
    \end{figure}

    Figure~\ref{fig:lut} shows an example of phase randomisation using LUTs. The \textit{pool} of random numbers available from the LUT is cycled through sequentially with successive frames and sub-frames continuing from where the previous left off.

	This paper presents a heuristic approach for investigating the effect of $N_{\text{LUT}}$ on edge enhancement and sub-frame independence. This is then tested on a binary phase ferroelectric device.

	\begin{figure}[h]
		\centering
		\begin{subfigure}[t]{0.16\textwidth}
			\includegraphics[width=\textwidth]{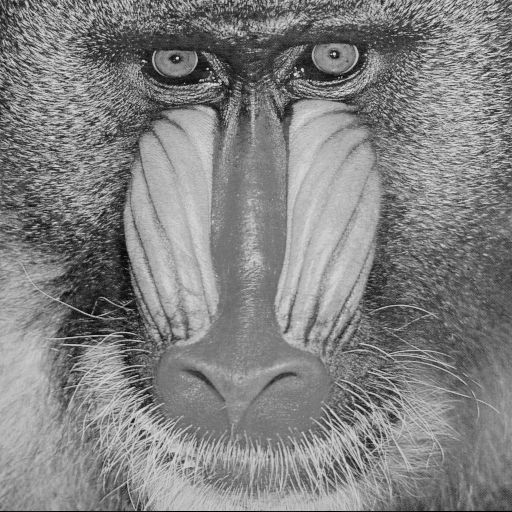}
			\label{fig:TestImage2}
		\end{subfigure}% DO NOT DELETE ME - It won't work properly without me
		\begin{subfigure}[t]{0.16\textwidth}
			\includegraphics[width=\textwidth]{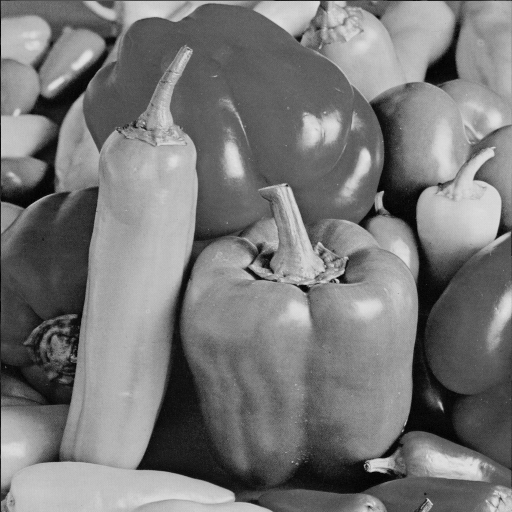}
			\label{fig:TestImage3}
		\end{subfigure}
		\begin{subfigure}[t]{0.16\textwidth}
			\includegraphics[width=\textwidth]{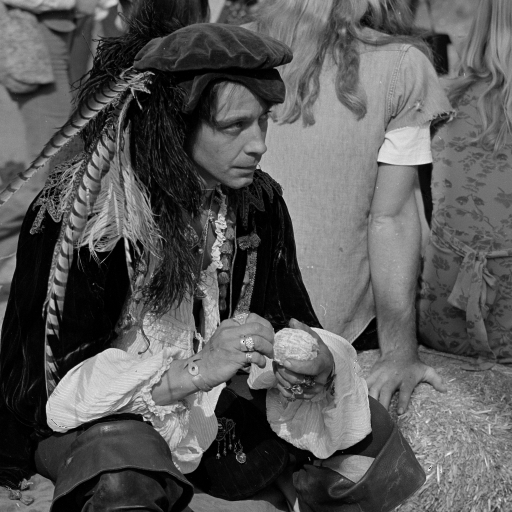}
			\label{fig:TestImage4}
		\end{subfigure}
		\begin{subfigure}[t]{0.16\textwidth}
			\includegraphics[width=\textwidth]{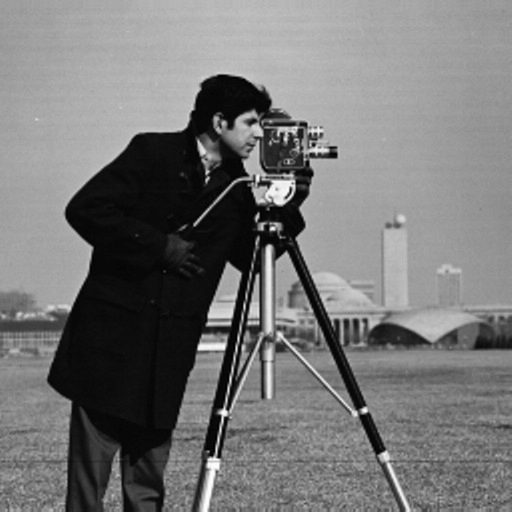}
			\label{fig:TestImage5}
		\end{subfigure}
		\begin{subfigure}[t]{0.16\textwidth}
			\includegraphics[width=\textwidth]{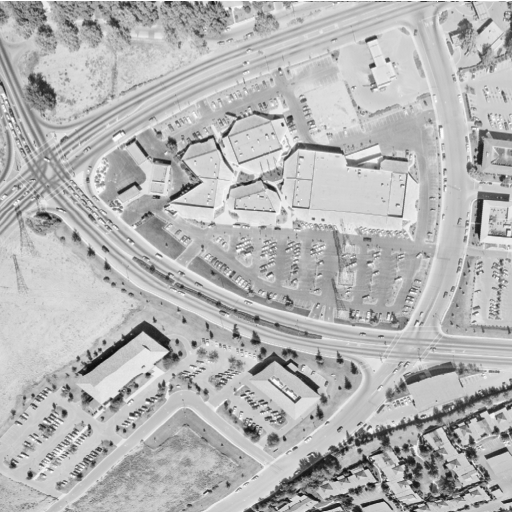}
			\label{fig:TestImage6}
		\end{subfigure}
		\begin{subfigure}[t]{0.162\textwidth}
			\includegraphics[width=\textwidth]{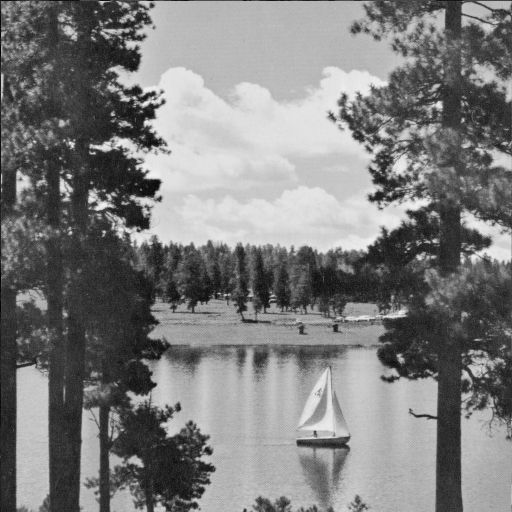}
			\label{fig:TestImage7}
		\end{subfigure}
		\caption{Standard test images from the USC-SIPI image database.\cite{sipidatabaseref} From left to right: Mandrill, Peppers, Man, CameraMan, Aerial, Landscape }
		\label{fig:TestImage}
	\end{figure}

	\subsection{Hard limits}
	
	For the sub-frame independence and edge enhancement reduction motivations, hard limits exist for theoretical correspondence between the LUT and psuedo-random number generator approaches.
	
	The hard limit for $N_{\text{LUT}}$ to meet the sub-frame independence constraint is that $N_{\text{LUT}}$ must be greater than $N_{\text{SF}}$, where $N_{\text{SF}}$ is the number of OSPR sub-frames. 
	
	To meet the edge enhancement constraint for both OSPR and GS, to ensure the hardware LUT approach is equivalent to the software pseudo-random number generator case, $N_{\text{LUT}}$ must be at least equal to $N_xN_yN_{\text{SF}}$ where $N_x$ and $N_y$ are the $x$ and $y$ resolutions respectively. This is impractical on many hardware devices so instead we attempt to determine the relationship between phase randomisation LUT length and final error.
	
	\subsection{Results}
    
\begin{figure}[h]
		\centering
		{\includegraphics[trim={0 0 0 0},width=0.7\linewidth,page=1]{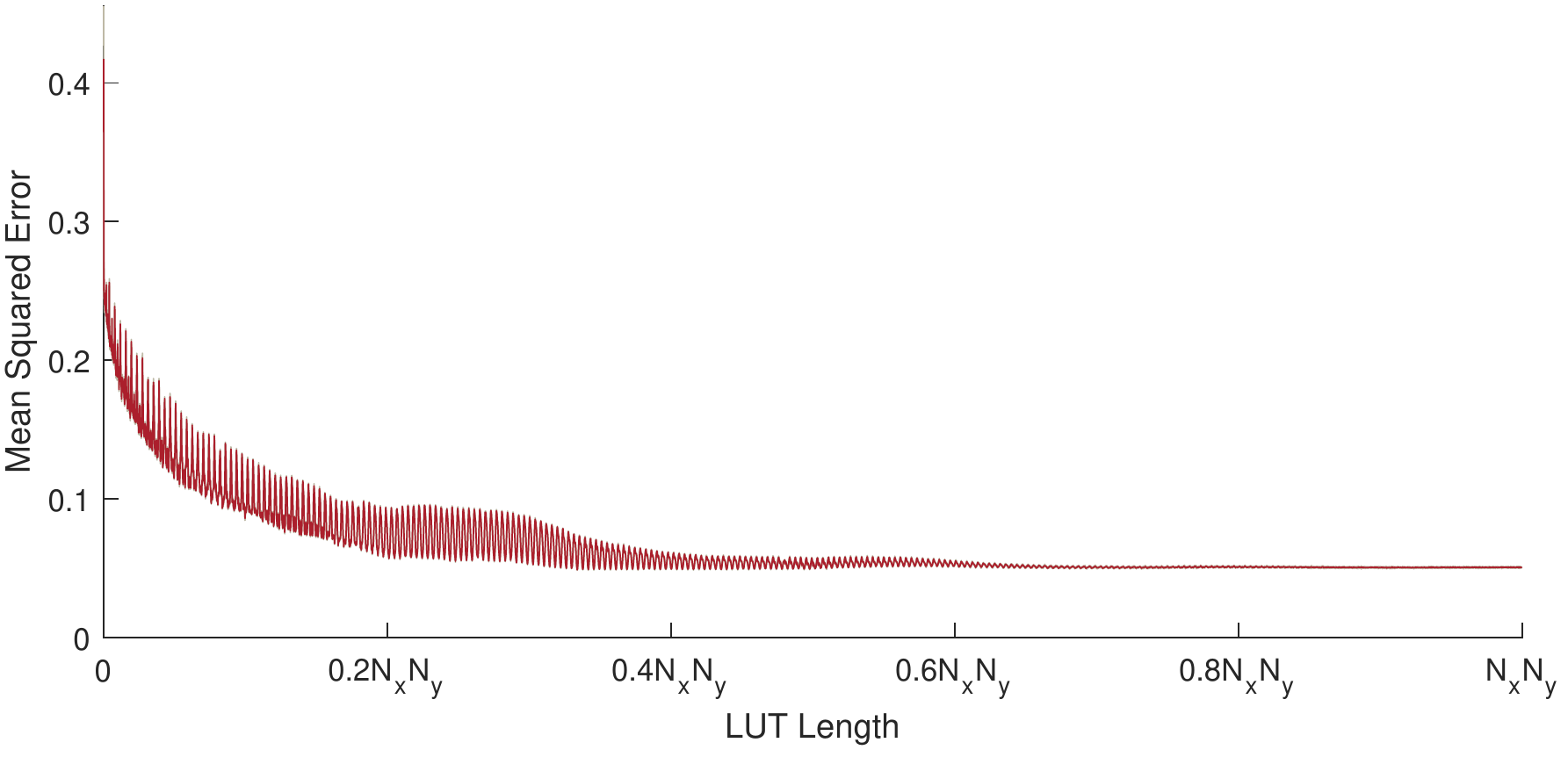}}
        {\includegraphics[trim={0 0 0 0},width=0.345\linewidth,page=1]{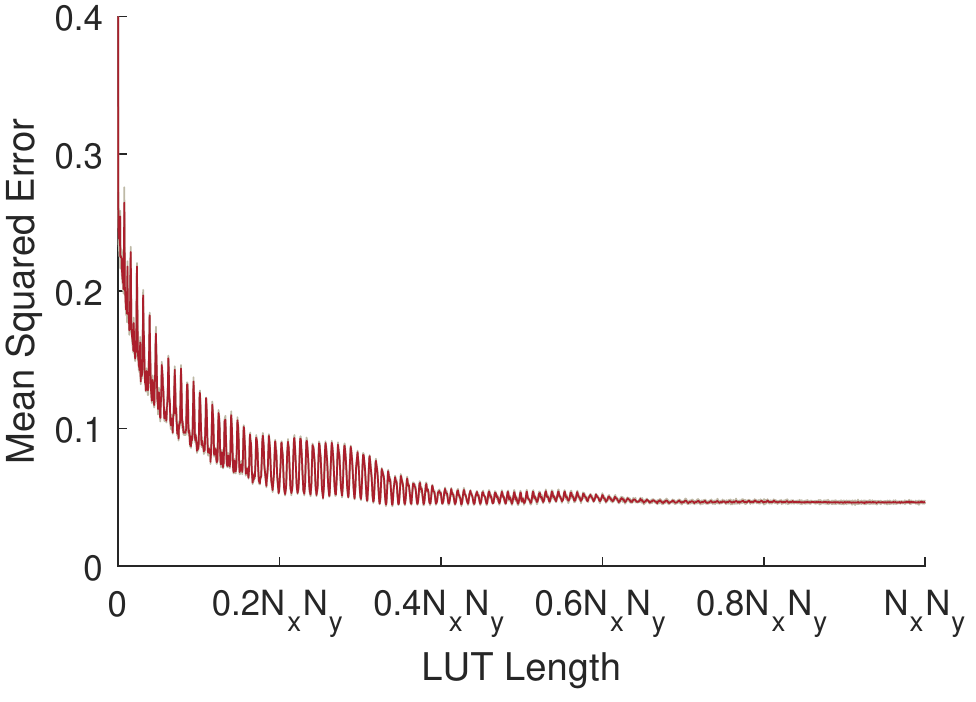}}
        {\includegraphics[trim={0 0 0 0},width=0.345\linewidth,page=1]{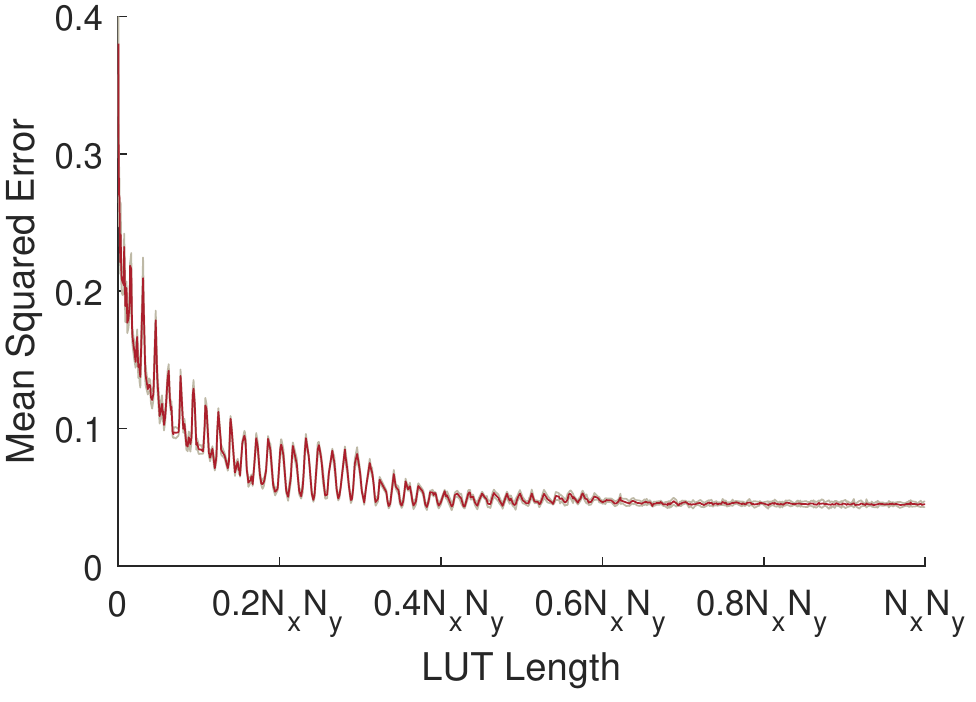}}
		\caption{Time averaged errors for prime values of $N_{\text{LUT}}$ for OSPR binary phase holograms with 24 sub-frames for the $6$ $256\times256$ (top) pixel test images shown in Figure~\ref{fig:TestImage}. Each value is the average of $100$ independent runs for each of the $6$ test images. The mean value is shown in red with two standard deviations above and below shown in grey. A comparison with $128\times128$ (bottom left) and $64\times64$ (bottom right) holograms is shown for comparison.}
		\label{fig:luts_001}
	\end{figure}

	Figure~\ref{fig:luts_001} shows the time averaged errors for prime values of $N_{\text{LUT}}$ for OSPR binary phase holograms with 24 sub-frames for the six $256\times256$ pixel test images shown in Figure~\ref{fig:TestImage}. A LUT for this would require over 12~Mb of space and realtime generation on a 60 FPS device would require multiple GFLOPs of processing power. Prime values of $N_{\text{LUT}}$ are used in order to avoid factorisation issues. Each error value shown is the average of $100$ independent runs and is normalised to the mean error for the \textit{Mandrill} test image to give the Normalised MSE~(NMSE). 
	
	\begin{align}
		NMSE_{image,n} = MSE_{image,n} \frac{MSE_{Mandrill,1000}}{MSE_{image,1000}}
	\end{align}
	
    The results show that, as expected, error is very high when $N_{\text{LUT}}$ is 0 and drops off rapidly until $N_{\text{LUT}}>N_{\text{SF}}$. Descent is slower there after with significant variation. Spikes occur at regular intervals when thedv cacc   cz LUT would be close in length to a multiple of the resolution. For example $N_{\text{LUT}}=256$ leads to a sharp increase in error as every image row has the same phase behaviour.
    
    Figure~\ref{fig:luts_001} also shows the effect of varying the hologram size $N_x\times N_y$. The similarity between the graphs suggests that for a given target error the required LUT size will be proportional to the total hologram size.
    
    \begin{figure}[h]
        \centering
        {\includegraphics[trim={0 0 0 0},width=0.7\linewidth,page=1]{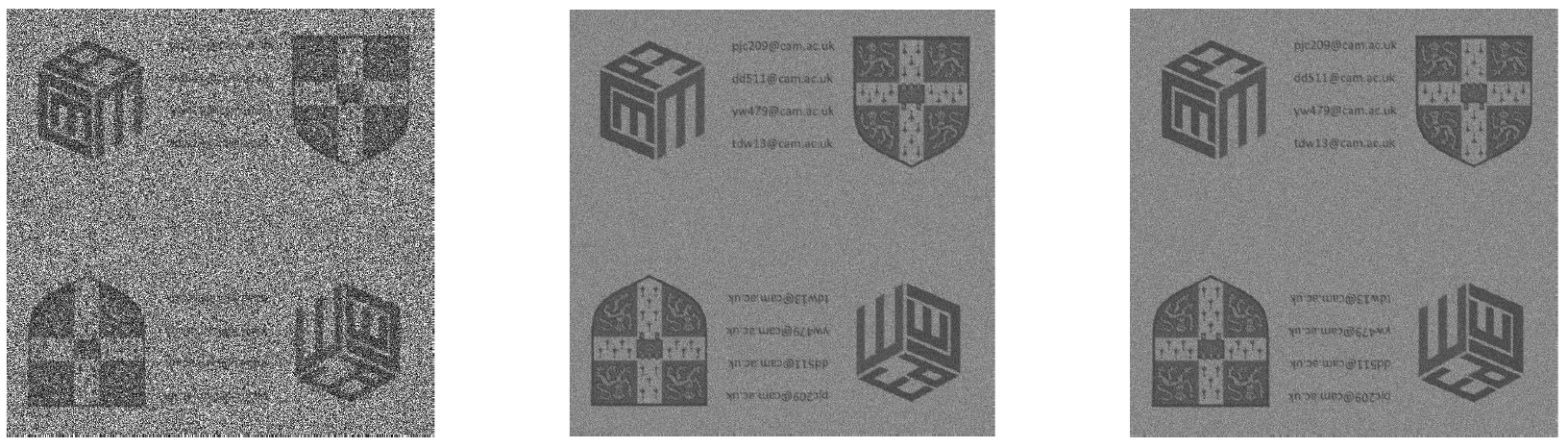}}
        \caption{Single OSPR sub-frame (left) with time average of 24 OSPR sub-frames with independent phase randomisation (centre) and $N_{\text{LUT}}=10007$ (right). The right hand image demonstrates less than $5\%$ additional error to the centre image.}
        \label{fig:luts_002}
    \end{figure}

    Figure~\ref{fig:luts_002} shows this effect on a real image. A single OSPR frame is shown right with 24 sub-frames in the centre. Both of these are with independent random numbers. The right hand image shown the case of $N_{\text{LUT}}=10007$, less than $0.03\%$ of the na\'\i ve length, and shows less than $5\%$ additional error.  The value of $10007$ for $N_{\text{LUT}}$ was chosen as the first prime number larger than $10000$ and similar results are seen for other large prime numbers. Independent trails with different test images showed no appreciable change in observed error. In practice the system admits of significant tuning for different test image resolutions as the oscillation in observed error is high in Figure~\ref{fig:luts_001}.

    \section{Validation}
    
    \begin{figure}[h]
        \centering
        {\includegraphics[trim={0 0 0 0},width=0.49\linewidth,page=1]{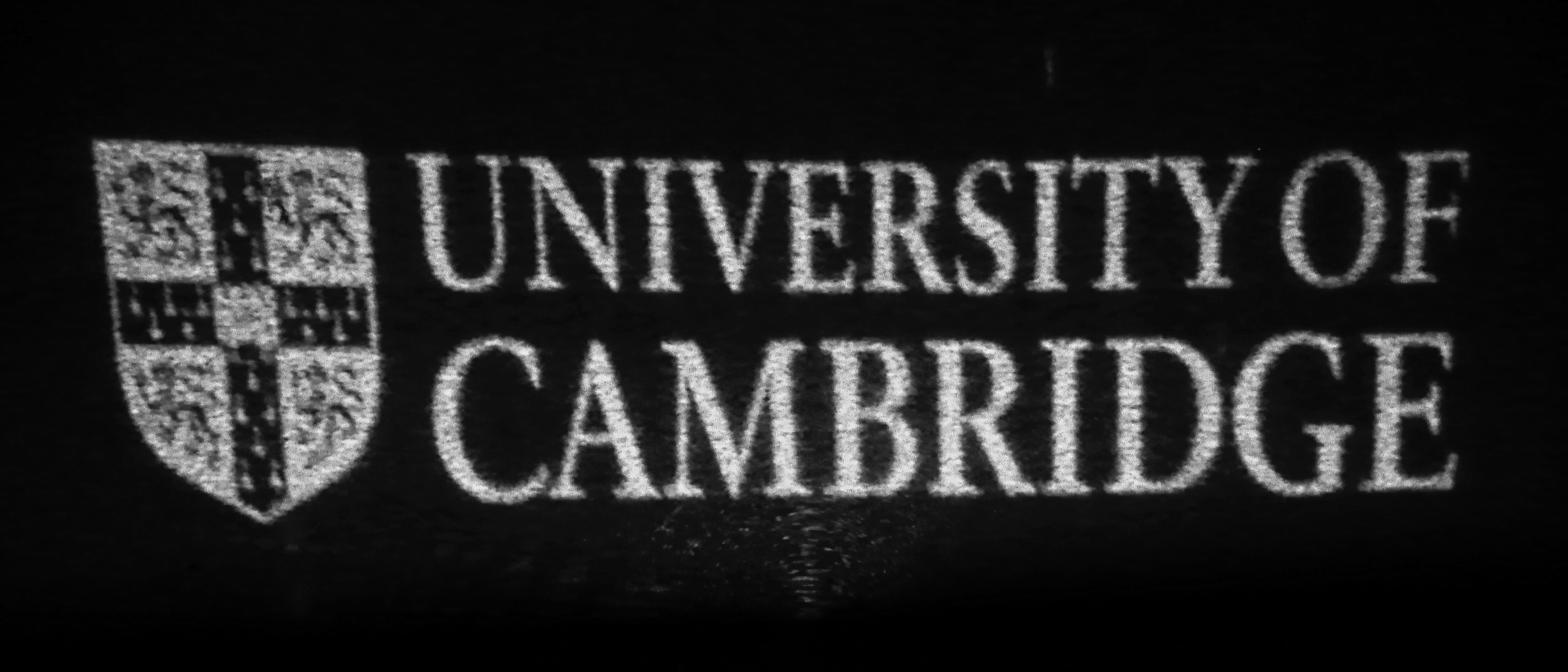}}
        {\includegraphics[trim={0 0 0 0},width=0.49\linewidth,page=1]{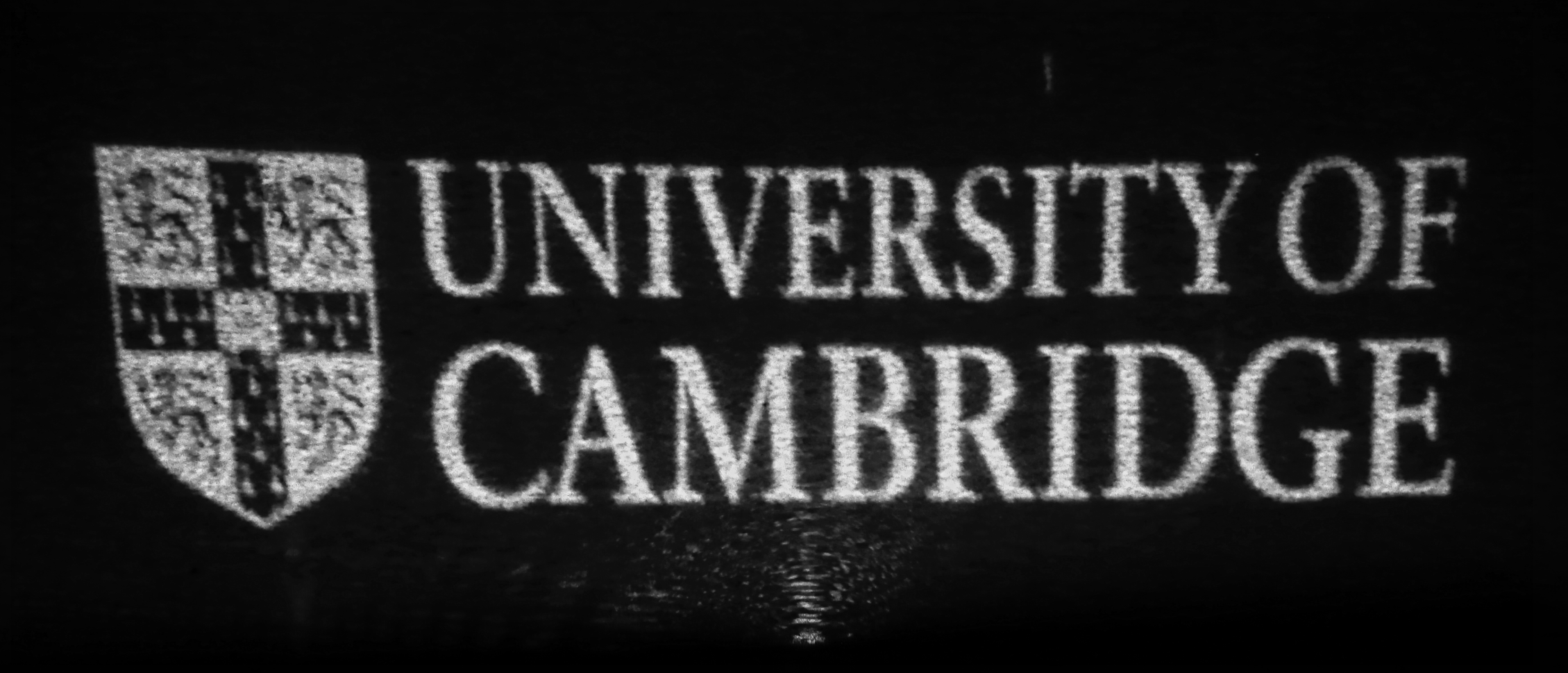}}
        \caption{Top portion of the replay field for 24 OSPR time-averaged sub-frames with independent phase randomisation (left) and $N_{\text{LUT}}=10007$ (right). Captured using a Canon 5D Mark III with a 24-105mm lens and a $\nicefrac{1}{60}$ second exposure.}
        \label{fig:luts_003}
    \end{figure}
    
    In order to validate our findings, we demonstrate the algorithm for a $1024\times1024$ binary ferroelectric display as shown in Figure~\ref{fig:luts_003}. The left hand image shows a fully independent randomisation while the right shows the case of $N_{\text{LUT}}=10007$, less than $0.03\%$ of the na\'\i ve length. The value of $10007$ for $N_{\text{LUT}}$ was chosen as the first prime number larger than $10000$ and similar results are seen for other large prime numbers. Independent trails with different test images showed no appreciable change in observed error.
    
    \section{Performance Discussion} 
    
    This paper has presented a simple method for removing the requirement for PRNGs and trigonometric functions from FPGA implementations of CGH. The process has the advantage of requiring only a single look-up operation.  As the LUT is read through cyclically, the memory location can be predetermined, improving performance at runtime. This performance benefit comes at the cost of reduced runtime flexibility and a larger LUT size. 
    
    For comparison, a fast PRNG approach such as M-sequences would offer a small algebraic overhead and still require the use of LUTs for the sine and cosine terms.
    
    For systems requiring runtime flexibility, PRNGs with LUTs for sin and cosine functions are expected to remain the dominant technique. For specific applications known at compile-time it is expected that this technique could offer performance and complexity benefits over existing approaches.

	\section{Conclusion}
	
    In this paper we initially reintroduced two reasons for phase randomisation in holographic projection: sub-frame independence and edge enhancement reduction. These have been discussed in the context of two algorithms: GS and OSPR. For embedded systems devices including FPGA and DSP these phase randomisation requirements can present a significant processing challenge and the use of LUTs can require many Mbs of storage for even small images
    
    As a result of this we have proposed that a LUT can be used. We also put forward three primary constraints on the minimum length of the LUT. First that the LUT have a prime length to reduce the chance of matching the periodicity of the image. Second that the LUT have a length greater than the number of sub-frames and third that the LUT should be greater than the largest dimension of the image. For the case of using a Gerchberg-Saxton~(GS) algorithm, only the first and third constraints apply.
    
    The proposed method was demonstrated using a LUT of length $10007$ for a simulated cases, less than $0.03\%$ of the length required for complete sub-frame independence. These both conformed to our expectations with the simulated image showing little less than $5\%$ additional error when using the LUT. Finally, we demonstrated this experimentally by generating frames for a binary phase projector. This also showed the expected behaviour with no visible quality differences when using a LUT of length $10007$.
    
    This result is of significance to embedded holographic systems as it negates the necessity for embedded complex number rotation operations and  reduces the required computation and the overall implementation complexity. This comes at the cost of system flexibility.
    
    \section*{Funding}
    
    The authors would like to thank the Engineering and Physical Sciences Research Council (EP/L016567/1, EP/T008369/1 and EP/L015455/1) for financial support during the period of this research.
    
    \section*{Disclosures}
    
    The authors declare no conflicts of interest.
        
    \bibliography{references}
	
\end{document}